\documentclass[a4paper]{article}
\usepackage{graphicx}  
\usepackage{slashed}
\usepackage{lscape}
\usepackage{amssymb}
\usepackage{mathtools}
\usepackage[margin=0.4cm]{caption}
\date{\today}
 \normalsize

\newcommand{\bmat}{\left(\begin{array}}
\newcommand{\emat}{\end{array}\right)}
\newcommand{\be}{\begin{equation}}
\newcommand{\ee}{\end{equation}}
\newcommand{\bea}{\begin{eqnarray}}
\newcommand{\eea}{\end{eqnarray}}

\def\gtwid{\mathrel{\raise.3ex\hbox{$>$\kern-.75em\lower1ex\hbox{$\sim$}}}}
\def\ltwid{\mathrel{\raise.3ex\hbox{$<$\kern-.75em\lower1ex\hbox{$\sim$}}}}
\def\gev{{\rm \, Ge\kern-0.125em V}}
\def\tev{{\rm \, Te\kern-0.125em V}}

\def    \be            {\begin{equation}}
\def    \ee            {\end{equation}}
\def    \bea           {\begin{eqnarray}}
\def    \eea           {\end{eqnarray}}

\def\d{\delta}

\def\gtwid{\mathrel{\raise.3ex\hbox{$>$\kern-.75em\lower1ex\hbox{$\sim$}}}}
\def\ltwid{\mathrel{\raise.3ex\hbox{$<$\kern-.75em\lower1ex\hbox{$\sim$}}}}
\def\gev{{\rm \, Ge\kern-0.125em V}}
\def\tev{{\rm \, Te\kern-0.125em V}}

\def    \be            {\begin{equation}}
\def    \ee            {\end{equation}}
\def    \bea           {\begin{eqnarray}}
\def    \eea           {\end{eqnarray}}

\def\d{\delta}

\def\d{\delta}

\def\beq{\begin{eqnarray}}
\def\eeq{\end{eqnarray}}

\begin{document}
\renewcommand{\thefootnote}{\fnsymbol{footnote}}
\vspace{.3cm}

\title{General Modified Friedmann Equations in Rainbow Flat Universe, by Thermodynamics}
\author{A. Ashour, M. Alcheikh  and N.Chamoun}

\author
{ \it \bf  A. Ashour$^{1}$\thanks{amaniamani.ashoor@gmail.com}, M. Alcheikh$^{1}$\thanks{mohammadalcheikh@yahoo.com} and  N.
Chamoun$^{2}$\thanks{nchamoun@ictp.it}
 \\
\small$^1$ Department of Mathematics, Faculty of Sciences, Damascus University, Damascus,
Syria.\\
\small$^2$  Physics Department, HIAST, P.O.Box 31983, Damascus,
Syria.
}

\maketitle

\begin{abstract}
{We investigate the derivation of Friedmann equations in Rainbow gravity following Jacobson thermodynamic approach. We do not restrict the rainbow functions to be constant as is customarily used, and show that the first law of thermodynamics with a corresponding `classical' proportionality between entropy and surface area, supplemented eventually by a `quantum' logarithmic correction, are not in general sufficient to obtain the equations in flat FRW metrics.  }
\end{abstract}


\begin{minipage}[h]{14.0cm}
\end{minipage}
\vskip 0.3cm \hrule \vskip 0.5cm

\section{Introduction}

Jacobson \cite{Jacobson} derived Einstein equations using a thermodynamic approach. As general relativity (GR) equations lead to Friedmann Equations (FEs), the approach was adopted \cite{Cai-Kim} in a cosmological setup to reach the FEs in general Friedman-Robertson-Walker (FRW) metrics.

This approach proved successful in investigating other modified gravities, such as $f(R)$ and scalar-tensor gravities \cite{Akbar-Cai} and Horava-Lifshtiz \cite{C-Ch-SH,Sheykhi}, where the FEs were reached albeit at the expense of adopting a proper definition of energy density and pressure. Recently, the authors of \cite{akram-sadat}, applied the Jacobson approach within Rainbow gravity, putting aside any `quantum' correction to the entropy-surface area relation, and reached the usual simple modified FEs (MFEs) mentioned in \cite{Mag-Smo}. However, the simple MFEs were obtained assuming the rainbow functions as constants. The general MFEs without this restriction were obtained first within rainbow gravity in \cite{Ling}, and our aim in this letter is to study these general MFEs using Jacobson approach.

Moreover, and rather than adopting an involved ``energy--supply vector'' technique, as is usually done, we use Kodama vector \cite{kodama} concept, which is a technique allowing a direct way to compute the energy flux through horizon.

\section{Jacobson approach and Friedamnn Equations}

We remind briefly how to derive FEs from thermodynamics within Jacobson approach. Our starting point is the FRW metric describing a homogenous and isotropic universe:
\begin{equation}
ds^2=-dt^2+a(t)(\frac{dr^2}{1-kr^2}+r^2d\Omega_{n-1} ^2) = h_{ab}dx^adx^b+\tilde{r}^2d\Omega_{n-1}^2, \label{1}
\end{equation}
where $d\Omega_{n-1}^2$ denotes the line element of an $(n-1)$-dimensional unit sphere \footnote{The volume of an $n$-dimensional unit ball is $\Omega_{n}=\pi^{n/2}/\Gamma(n/2+1)$.}, $k=+1,0$ and $-1$ for a closed, flat and open universe respectively, $\tilde{r}=a(t)r$ and the indices $a,b$ span ${0 (t),1 (r)}$.

The dynamical apparent horizon is determined by solving $h^{ab}\partial_{a}\tilde{r}\partial_{b}\tilde{r}=0$ to obtain its radius given by:
\begin{equation}
\tilde{r}_A=\frac{1}{\sqrt{H^2+k/a^2}}, \label{3}
\end{equation}
where  $H$ is the Hubble parameter, $H\equiv\dot{a}/a$ \footnote{This radius is identical to the Hubble horizon radius $\tilde{r}_H=1/H$ for flat universe ($k=0)$}.
The apparent horizon is a causal horizon associated with the gravitational entropy and surface gravity \cite{Bak} onto which one can apply the first law of thermodynamics. If we define   the total energy inside the space with radius $\tilde{r}$  by
\begin{equation}
E=\frac{n(n-1)}{16\pi G}\tilde{r}^{n-2}(1-h^{ab}\partial_a\tilde{r}\partial_b\tilde{r}), \label{8}
\end{equation}
then the energy conservation through the apparent horizon is expressed as:
\begin{equation}
\nabla_a E= A \Psi_a+W\nabla_a V, \label{7}
\end{equation}
where $A=n\Omega_{n}\tilde{r}^{n-1}$ ($V=\Omega_{n}\tilde{r}^{n}$) is the the area (volume) of the apparent horizon, and $W, \Psi_a$ are the work density and the energy-supply vector at the apparent horizon given by
\bea W=-\frac{1}{2}T^{ab}h_{ab} &,&  \Psi_a={T_{a}}^b\partial_{b}\tilde{r}+W\partial_{a}\tilde{r} \label{5} \eea
with $T^{ab}$ the 2-dim projection of the $(n+1)$-dimensional energy-momentum tensor $T^{\mu\nu}$ of a perfect fluid matter in the normal direction to the $(n-1)$-sphere.

The first law of thermodynamics relates the change of energy inside the apparent horizon due to the heat flux, expressed by the first term of Eq. \ref{7}, to the entropy via
\bea
dE &=& -\delta{Q}=TdS, \label{firstlaw}
\eea
Assuming a perfect fluid matter for the universe, we have $T_{\mu\nu}=(\rho+P)U_\mu U_\nu+Pg_{\mu\nu}$ with $(U_\mu, \rho, P)$ denote respectively the four-velocity, the energy density and pressure of the fluid given usually via the equation of state $(P=\omega \rho)$. One gets
\begin{equation}
\Psi_a=(-\frac{1}{2}H\tilde{r}(\rho+P),\frac{1}{2}(\rho+P)a), \label{11}
\end{equation}
Assuming the apparent
horizon has an entropy $S$ and temperature $T$ given by \cite{Hawk}
\begin{equation}
S=\frac{A}{4G},
\qquad
T=\frac{1}{2\pi\tilde{r_A}}, \label{13}
\end{equation}
then Eq. (\ref{firstlaw}) gives one FE:
\begin{equation}
\dot{H}-\frac{k}{a^2}=-\frac{8\pi G}{n-1}(\rho+P), \label{14}
\end{equation}
Substituting $(\rho+P)$ from the conservation equation
\begin{equation}
\dot{\rho}+nH(\rho+P)=0, \label{16}
\end{equation}
and integrating, we get another FE expressing the $00$ component of Einstein equations:
\begin{equation}
H^2+\frac{k}{a^2}=\frac{16\pi G}{n(n-1)}\rho, \label{17}
\end{equation}

\section{Friedmann equation in rainbow gravity from the first law of thermodynamics}

The metric in rainbow gravity formalism is given by \cite{Mag-Smo}:
\begin{equation}
ds^2=-\frac{1}{f^2(\varepsilon)}dt^2+\frac{a^2}{g^2(\varepsilon)}(\frac{1}{1-kr^2} dr^2+r^2d\Omega^2_{n-1}), \label{20}
\end{equation}
The original work \cite{Mag-Smo} deduced the MFEs in the case where the rainbow functions $f(\varepsilon),g(\varepsilon)$ were constant. Nonetheless, it was argued \cite{Ling} that while in considering a specific measurement the probe energy $\varepsilon$ appearing in the metric can be treated as a constant independent of spacetime coordinates, the evolution with cosmological time of this very energy $\varepsilon$,  when identified with that of massless particles such as photons or gravitons, should be taken into account during a longtime process studying the semi-classical effects of particles on the background metric. Thus, in the general case, the rainbow functions $f(\varepsilon (t)),g(\varepsilon (t))$ should be treated as implicitly time dependent, with the condition:
\begin{equation}
\lim_{\varepsilon/\varepsilon_P \to 0}  f=1, \qquad \lim_{\varepsilon/\varepsilon_P \to 0}g=1, \label{20b}
\end{equation}
where $\varepsilon_P$ is the Planck energy.

The general MFEs  in the case of flat FRW universe (k=0) and general rainbow functions were first deduced in \cite{Ling} using modified Einstein equations and found to be:
\bea
(H-\frac{\dot{g}}{g})^2&=&\frac{8\pi G}{3f^2}\rho, \label{30}
\\
\dot{H}-\frac{\ddot{g}}{g}+\frac{\dot{g}^2}{g^2}&=&-\frac{4\pi G}{f^2}(\rho+P)-(H-\frac{\dot{g}}{g})\frac{\dot{f}}{f}, \label{31}
\eea

While the work of \cite{akram-sadat} deduced the simple form of Eqs. (\ref{30},\ref{31}) in the case of $\dot{f}=\dot{g}=0$, we shall apply the Jacobson approach in the general case and investigate when it can impose the general MFEs.

Rewriting the metric (Eq. \ref{20}) with $\tilde{r}=(a/g)r$ as
\bea
ds^2=h_{ab}dx^adx^b+\tilde{r}^2d\Omega^2_{n-1} \label{21} &,& h_{ab}=diag(-1/f^2,a^2/g^2),
\eea
 we  solve the relation $h^{ab}\partial_{a}\tilde{r}\partial_{b}\tilde{r}=0$ and determine the radius of the apparent horizon
\bea
\tilde{r}_A=\frac{g}{fH}, \label{22} &,&
\dot{\tilde{r}}_A=-\frac{f^2}{g^2}\tilde{r}_A^3H[(\frac{\dot{f}}{f}-\frac{\dot{g}}{g})H+\dot{H}], \label{23}
\eea

In order to compute the heat flux, we use the Kodama approach \cite{kodama} in $(1+3)$-dim to calculate the energy of particles tunneling through the apparent horizon, where the Kodama vector:
\bea
K^a=-\epsilon^{ab}\nabla_b \tilde{r} &,& \epsilon_{ab} = g_{rr}(dt)_a\wedge(dr)_b \label{Kvector definition}
\eea
is related to the 4-momentum flow $J_a$ and the heat flux $\d Q$ through the horizon via
\bea
J_a=T_{ab}K^b, \label{momentum}
\delta Q=\int_H J_ad\Sigma^b=\int_H T_{ab} K^a d\Sigma^b
\eea
 The Kodama vector is very similar to the
Killing vector $(\partial/\partial t)^a$ in the de Sitter space. In the stationary black hole spacetime, the
timelike Killing vector is used to define a conserved energy, but as there is
no timelike Killing vector in FRW spacetime, the Kodama
vector generates a preferred flow of time and is a dynamic analogue of a stationary Killing
vector \cite{Hay1}.

In our case, using Eqs (\ref{21},\ref{Kvector definition},\ref{momentum}) we get:
\bea
K^a&=&-\frac{1}{g^2} \lbrack-(\frac{\partial}{\partial t})^a+Hr(\frac{\partial}{\partial r})^a \rbrack, \label{Kvector}
\\
\delta Q&=&AK^aT_{ab}n^b\vert_{\tilde{r}=\tilde{r}_A}
=A(K^tT_{tt}n^t+K^rT_{rr}n^r)\vert_{\tilde{r}=\tilde{r}_A}
=\frac{1}{g^2}A(\rho+P)H\tilde{r}_A, \label{25}
\eea
with the generator normal vector of horizon given by  $n_a = (\partial/\partial t)^a - Hr (\partial/\partial r)^a$, and the components of the energy momentum tensor are given by
\bea
T_{tt}=\rho f^{-2} &,&
T_{rr}=(a/g)^2P
\eea
We stress here that
the relation connecting the horizon temperature with the
geometry of the universe depends only on this latter
and not on the specific theory of gravity under study \cite{saridakis}. Thus, the RHS of Eqs (\ref{13}) should remain the same albeit with a different radius $\tilde{r_A}$.
However, for the LHS of Eqs (\ref{13}) , it is well known \cite{cai, saridakis} that the `classical' linear relation between entropy and surface-area may change, and in many theories of quantum gravity, logarithmic leading order 'quantum' corrections may arise \footnote{In \cite{1003.0876}, a new form of `non-additive' entropy, suitable for gravitational systems, was used. However, we shall not follow this approach here, and will be content with the Bekenstein-Hawking entropy}. In the context of rainbow gravity, studies were performed \cite{galan, zhang} and give the result
\bea
\label{gen-entr} S &=& \frac{A}{4G} - \alpha \ln \frac{A}{4G}
\eea
where $\alpha = 1/2$ in \cite{zhang}.
 Moreover, the entropy $S$ is in its turn dependent on the Probe particle's energy via the dependence of the  surface $A$ on  $\tilde{r}_A$ (Eq. \ref{23}).
Substituting in the Clausius relation (Eq. \ref{firstlaw}) and using Eq. (\ref{23}) we get
\begin{equation}
\frac{\dot{g}}{g}H-\frac{\dot{f}}{f}H-\dot{H}=\frac{4\pi G}{f^2}\left(1-\frac{\alpha G}{\pi \tilde{r_A}^2}\right)^{-1}(\rho+P), \label{27}
\end{equation}
The modified conservation equation in rainbow gravity given by \cite{Awad} \footnote{We are assuming that no cosmological constant $\Lambda$, and that Newton's constant $G$ does not vary with time. Moreover, the parameter $\omega$ in the equation of state $P=\omega \rho$ is also assumed constant.}
\begin{equation}
\dot{\rho}+3(H-\frac{\dot{g}}{g})(\rho+P)=0, \label{24}
\end{equation}
allows to compute $\rho+P$, which when substituted in Eq. (\ref{27}) leads to
\begin{equation}
f^2(H-\frac{\dot{g}}{g})[\dot{H}+\frac{\dot{f}}{f}H-\frac{\dot{g}}{g}H] \left( 1-\alpha \frac{G}{\pi \tilde{r_A}^2}\right) =\frac{4\pi G}{3}\dot{\rho},
\end{equation}

Using Eq.(\ref{23}), we see here that provided we impose the constraint
\begin{equation} \label{gen-constraint} \left[
\frac{\ddot{g}}{g}+\frac{\dot{f}}{f}\frac{\dot{g}}{g}-\frac{\dot{g}}{g}H-\frac{\dot{g}^2}{g^2}\right] - \alpha \left[\frac{G}{\pi} \frac{H^3f^2}{g^2} \left(\frac{\dot{f}}{f}-\frac{\dot{g}}{g}+\frac{\dot{H}}{H} \right)\right] = 0
\end{equation}
we get
\begin{equation}
f^2(H-\frac{\dot{g}}{g})\left(\dot{H}-\frac{d}{dt}(\frac{\dot{g}}{g})\right)+f\dot{f}(H-\frac{\dot{g}}{g})^2=\frac{4\pi G}{3}\dot{\rho}, \label{28}
\end{equation}
which upon integration leads to the first MFE (Eq. \ref{30}).

To get the second MFE (Eq. \ref{31}), it suffices to add the constraint Eq.(\ref{gen-constraint}) to (\ref{27}). We conclude thus that the Jacobson approach cannot  lead to the MFEs in Rainbow gravity (Eqs. \ref{30},\ref{31}) unless the constraint (Eq. \ref{gen-constraint}) is met.

\section{Discussion}
We discuss now the physical implications of Eq.(\ref{gen-constraint}). Clearly, when $g$ is a constant, the `classical' part of the constraint is satisfied whereas the `quantum' part is not met unless $f$ is inversely proportional to $H$, but the question arises as to whether other solutions are of physical significance. In order to simplify the discussion and get exact solutions, we shall assume that the universe apparent horizon $\tilde{r_A}$ is far larger than the Planck length $L_P \propto \sqrt {G}$, which is equivalent to looking at times far beyond the Planck time. In this regime, one can neglect the quantum corrections in the entropy-area relation, proportional to $\alpha$, and so we assume the constraint of the form:

\bea
\label{299}
\frac{\ddot{g}}{g}+\frac{\dot{f}}{f}\frac{\dot{g}}{g}-\frac{\dot{g}}{g}H-\frac{\dot{g}^2}{g^2} &=& 0
\eea

Actually, one can verify that the MFEs (Eqs. \ref{30} \& \ref{31}) are equivalent to (Eqs. \ref{30} \& \ref{24}). Originally, Jacobson approach proved in the context of GR that the first law of thermodynamics (Eq. \ref{firstlaw}) is equivalent to one FE (Eq. \ref{14}), and that when supplemented with energy conservation (Eq. \ref{16}) one recovers the two FEs. However, in generalized rainbow gravity, we have just proved that, provided the constraint Eq. (\ref{299}) is met and only then, the first law expressed by (Eq. \ref{27}) supplemented  with the energy conservation (Eq. \ref{24}) are equivalent to the two MFEs, in which case one can trade, when energy conservation is respected, the first law of thermodynamics with the equation of motion. In rainbow gravity with $\dot{f}=\dot{g}=0$, thermodynamics first law and energy conservation do replace the equations of motion, a situation which is not valid in generalized rainbow gravity when the constraint is not met. This constraint leads then, when one insists on the validity of first law of thermodynamics in the sense developed earlier, that the rainbow functions $(f,g)$ are not independent but are related.

In principle, if we know the temporal laws $f(t),g(t)$, then using Eqs. (\ref{30},\ref{24}), we reach an ODE:
\bea \dot{\rho} + 3 c_1 \frac{\rho^{3/2}}{f(t)}(1+\omega) &=&0 \label{ODE}\eea with $c_1=\sqrt{\frac{8\pi G}{3}}$, which can be solved to get consecutively $\rho(t)$, $H(t)$, then  the temporal evolution of the scale factor $a(t)$. However, we are generally given $f(\varepsilon), g(\varepsilon)$, and unless one knows the dependence of the probe energy on time ($\varepsilon(t)$), we can not solve the equations of motion. In ordinary rainbow gravity, the probes are carried during cosmologically short periods of time when one can assume the probe energy $\varepsilon(t)$ approximately constant. One exception, where one can do the computations when the probe energy varies with time, lies in the arena of radiation-dominated early universe, where the probe particles can be taken as photons, and on  dimensional grounds we have \cite{lingwu} \bea \label{varepsilon} \varepsilon = \bar{\varepsilon} &\propto&  \rho^{1/4}\eea
Then we reach again the ODE (\ref{ODE}) with $f$ now as a function of $\rho$, and by solving it we get consecutively ($\rho(t), f(t), g(t), H(t)$), then ($a(t)$) can be found.

As an example, we take the case of $f=1$, then we get an ODE $\dot{\rho}+K \rho^{3/2}=0$ with $K=3c_1(1+\omega)$ with solution assuming singularity, when extrapolated into the Big Bang, given by
\bea \rho=\frac{4}{K^2 t^2} &,& t=\frac{2}{K \varepsilon^2 } \label{example}\eea
Many solutions existing in the literature \cite{lingwu} though do not satisfy (Eq. \ref{firstlaw}), as they do not comply with the constraint (Eq. \ref{299}). Imposing the constraint would make the rainbow functions related to each other. For this let us assume we are given $f(\varepsilon)$, and seek how to determine $g(\varepsilon)$. The first strategy consists of overlooking the thermodynamical constraint (Eq. \ref{299}) and take any acceptable function $g(t)$. One then solves for $\rho(t), H(t)$, using Eqs. (\ref{30}, \ref{24}), but the obtained solution may not satisfy thermodynamics first law even though it might be physically meaningful regarding other considerations. If we return to the previous example ($f=1$), and if we take, say, $g(t)=t$, then we find that we have $H=\frac{\gamma}{t}$ with $\gamma=1+ \frac{2/3}{1+\omega}>0$, giving a power law for the expansion of the universe ($a(t)\propto t^\gamma$). However, clearly this choice of functions does not satisfy the thermodynamic constraint, as Eq. (\ref{299}) would give instead $H=-\frac{1}{t}$. Thus, we adopt the second strategy which consists of imposing Eq. (\ref{299}) and solving for $g$ by replacing $H$ in the constraint by $H=\frac{\dot{g}}{g}+\frac{c_1 \rho^{1/2}}{f}$ from Eq. (\ref{30}). Going back to our example where $f=1$, we find an ODE for $g(t)$ to be solved and then using Eq. (\ref{example}) we get $g(\varepsilon)$: \bea g \ddot{g} - 2 \dot{g}^2 -(\gamma-1) g \dot{g} =0 &\Rightarrow& g(\varepsilon) = \frac{K_2 \varepsilon^{2 \gamma}}{1- \gamma K_1 \varepsilon^{2 \gamma}}\eea where $K_{1,2}$ are constants. Knowing $f(t),g(t)$, one can advance now to compute consecutively $\rho(t), H(t), a(t)$ while being assured the found solution will satisfy thermodynamics requirements.

One can proceed differently, in that we, up till this point, were given the rainbow function as a starting point followed by calculating energy density and Hubble constant. We assume now $\rho(t)$ as given, and seek acceptable rainbow functions $f,g$ which meet the thermodynamic constraint Eq. (\ref{299}). Let's discuss the important case of a constant energy density ($\dot{\rho}=0$), which in ordinary GR corresponds to inflationary scenario. Imposing Eq. (\ref{24}), we find either ($H-\frac{\dot{g}}{g}=0$) leading via eq. (\ref{30}) to the uninteresting result of zero energy density, or $\rho+P=0$, so we find here ($\omega=-1$) exactly like in GR situation. Now the thermodynamics first law via Eq. (\ref{27}) and the constraint (\ref{299}) would give
\bea  H=c_2\frac{g}{f} &,& f \ddot{g} + \dot{f} \dot{g} -c_2 \dot{g} g - \frac{f \dot{g}^2}{g}=0\label{eq}\eea where $c_2$ is an integration constant.
We still have two degrees of freedom represented by the functions ($f,g$). If we take the first ansatz $f=1$ then the solution of the ODE representing the constraint is of the form
\bea g=\frac{c_3 e^{c_3 t}}{1-c_2 e^{c_3 t}} &\Rightarrow& a(t) = \frac{c_4}{1-c_2 e^{c_3 t}}\eea where $c_{3,4}$ are again integration constants. However, this solution does not represent inflation.

 If we take now the other ansatz of proportional rainbow functions $f=\delta g$, then $H=\frac{c_2}{\delta}=h$ constant, and we have inflation in the form $a(t) \propto e^{h t}$. To find the rainbow function, the constraint gives now $\ddot{g}=h\dot{g}$ and so we get the solution
 \bea \frac{f}{\delta}=g(t)&=& c_5 e^{ht} - \frac{c_6}{h}\eea ($c_i$'s are constants). One can thus conceive a situation where $g \ll 1$ at the start of inflation, then $g$ increases till it reaches the value $1$ approximately at the end of inflation.

 We summarize our findings in that imposing the validity of deducing the equations of motion by applying Thermodynamics first law in the context of generalized rainbow gravity makes the generalized rainbow functions dependent on each other.

\end{document}